\documentclass[12pt,a4paper]{article}
\usepackage[utf8]{inputenc}
\usepackage{latexsym,amssymb,euscript}
\usepackage{graphicx}
\usepackage[numbers,sort&compress]{natbib}
\usepackage{amsmath}
\usepackage{nicefrac}
\usepackage{booktabs}
\usepackage[linktocpage]{hyperref}
\usepackage{chngcntr}
\usepackage{ragged2e}
\usepackage{array}
\usepackage{caption}
\usepackage{subcaption}
\usepackage[titletoc]{appendix}
\usepackage[normalem]{ulem}
\graphicspath{{./figures/}}
\hypersetup{
 linktocpage = false,
 urlcolor = urlblue,
 colorlinks = true,
 linkcolor = urlblue,
 anchorcolor = citegreen,
 citecolor = urlblue,
 pdfstartview = {XYZ null null 1.25}
           }
\usepackage[left=2cm, right=2cm]{geometry}
\usepackage[section]{placeins}
\usepackage{color}
\usepackage{xcolor}
\definecolor{urlblue}{rgb}{0.2,0.4,0.7}
\definecolor{citegreen}{rgb}{0,0.4,0.2}
\definecolor{linkred}{rgb}{0.9,0.2,0.1}
\usepackage{float}
\usepackage{academicons}
\definecolor{orcidlogocol}{HTML}{A6CE39}
\usepackage{changepage}
\usepackage{fancyhdr}
\newcommand{\drv}{{\rm d}}
\newcommand{\as}{\alpha_s}
\newcommand{\abar}{\bar{\alpha}_s}
\newcommand{\LQCD}{\Lambda_{\rm QCD}}
\newcommand{\MSb}{\overline{\rm MS}}
\newcommand{\LO}{{\rm LO}}
\newcommand{\NLO}{{\rm NLO}}

\newcommand{\BLM}{{\rm BLM}}

\newcommand{\shat}{\hat{s}}
\newcommand{\yIP}{y_{I\!P}}
\newcommand{\cso}{c_{s_0}}
\newcommand{\muR}{\mu_R}

\newcommand{\orcidJVB}{\href{https://orcid.org/0000-0002-3917-306X}{\includegraphics[scale=0.1]{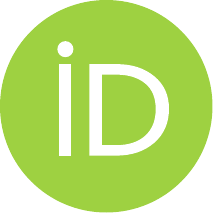}}}
\newcommand{\orcidVPG}{\href{https://orcid.org/0000-0003-4943-9973}{\includegraphics[scale=0.1]{logo-orcid.pdf}}}

\renewcommand{\arraystretch}{1.6}

\begin{document}

\begin{titlepage}

{
\begin{adjustwidth}{-1cm}{-1cm}
\begin{center}
  {\LARGE \bf Diffractive Photoproduction of High $p_T$  Photons\\[0.3cm]  in Ultraperipheral Collisions at the LHC\\[0.3cm]}
\end{center}
\end{adjustwidth}
}

\vskip 0.75cm

% \WIP{author list to be confirmed}
\centerline{
Jo\~ao~V.~Bulh\~oes,$^{\,1\dagger}${\orcidJVB}
and
Victor~P.~Gon\c{c}alves$^{\,1\ddagger}${\orcidVPG}
}

\vskip .4cm

\centerline{${}^1$ {\sl Institute of Physics and Mathematics, Federal University of Pelotas, 354, 96010-900, Pelotas, RS, Brazil}}
\vskip 1.15cm

\begin{abstract}
\vspace{0.25cm}
\headrule
\vspace{0.50cm}

We investigate diffractive photoproduction of large-transverse-momentum photons in ultraperipheral collisions at the LHC, considering different collision systems and alternative descriptions of the BFKL Pomeron. We estimate cross sections and differential distributions including NLO corrections to the BFKL characteristic function and assess the prospects for measuring this process in ultraperipheral pp, pPb, and PbPb collisions. We find that the cross sections are sizable, and that future measurements in this channel can help constrain BFKL dynamics and improve our understanding of high-energy QCD.

\vspace{0.50cm}
\headrule
\vspace{0.75cm}
{
 \setlength{\parindent}{0pt}
 \textsc{Keywords}: \vspace{0.15cm} \\
 High-energy QCD, 
 BFKL resummation, 
 Ultraperipheral collisions, 
 Diffractive photon production.
\vspace{0.65cm}
\headrule
}
\end{abstract}
\vspace{-0.00cm}
\vfill
$^{\dagger}${\it e-mail}:
\href{mailto:joaovitor1729@gmail.com}{joaovitor1729@gmail.com}
\\
$^{\ddagger}${\it e-mail}:
\href{mailto:barros@ufpel.edu.br}{barros@ufpel.edu.br}

\end{titlepage}

\tableofcontents
\clearpage

%-----------------------------------------
\section{Introduction}
\label{sec:intro}
%-----------------------------------------
Describing strong interaction dynamics at high energies remains a significant theoretical challenge within the Standard Model of particle physics. While the theoretical treatment of Quantum Chromodynamics (QCD) at small Bjorken $x$ has advanced considerably over the last few decades, several questions remain unresolved. This ongoing challenge continues to motivate the analysis of diverse final states and the proposal of next-generation colliders, which are essential for refining our understanding of hadronic structure.

Studies of $ep$ collisions at HERA, specifically processes such as $ep \rightarrow e YX$, where $Y$ and $X$ represent the hadronic systems of the dissociated photon and proton, have demonstrated that analyzing diffractive processes is a promising method for constraining QCD dynamics (for a  review, see, e.g., Ref.~\cite{Schoeffel:2009aa}). In particular, the analysis of processes involving final states $X$ and $Y$ separated by a large rapidity gap has been proposed as a probe of the Balitsky–Fadin–Kuraev–Lipatov (BFKL) Pomeron~\cite{Fadin:1975cb,Kuraev:1976ge,Kuraev:1977fs,Balitsky:1978ic}. Existing measurements of double dissociation, rapidity gaps between jets, and diffractive vector-meson production at large   momentum transfer $t$ pose theoretical challenges~\cite{Schoeffel:2009aa}. In the first two cases, additional strong interactions between hadronic systems can destroy the rapidity gap generated by diffractive hard scattering. While diffractive vector-meson production ($\gamma p \to V X$) avoids this issue, it relies on knowledge of the vector-meson wave function (see, e.g., Refs.~\cite{Enberg:2003jw,Poludniowski:2003yk}). Conversely, diffractive photon production ($\gamma p \to \gamma X$) circumvents both gap-destruction effects and wave-function uncertainties: the $Y$ system is a single photon, preventing subsequent strong interactions with the proton remnant, and the $\gamma \to q\bar q$ splitting occurs electromagnetically. A hard scale is established by the large transverse momentum of the scattered photon, as $p_T^2 \approx -t$ in photoproduction, where the large momentum transfer also induces proton dissociation into the $X$ system. The hard scattering is thus calculable in perturbative QCD and, apart from the BFKL energy scale, which is fixed by data, the main non-perturbative input is provided by the proton PDFs. These characteristics have motivated earlier studies that considered the leading-order (LO) solution of the BFKL equation \cite{Ginzburg:1985tp,Ginzburg:1996vq,Ivanov:1998jw,Cox:1999kv,Evanson:1999zb}. The impact of next-to-leading-order (NLO) corrections on diffractive photon production at high-$t$ was subsequently estimated in Ref.~\cite{Goncalves:2013pac}. Unfortunately, the scarcity of HERA data and the decommissioning of the $ep$ collider have limited the ability to derive more precise conclusions.  

\begin{figure}[t]
\centering
\includegraphics[width=0.65\textwidth]{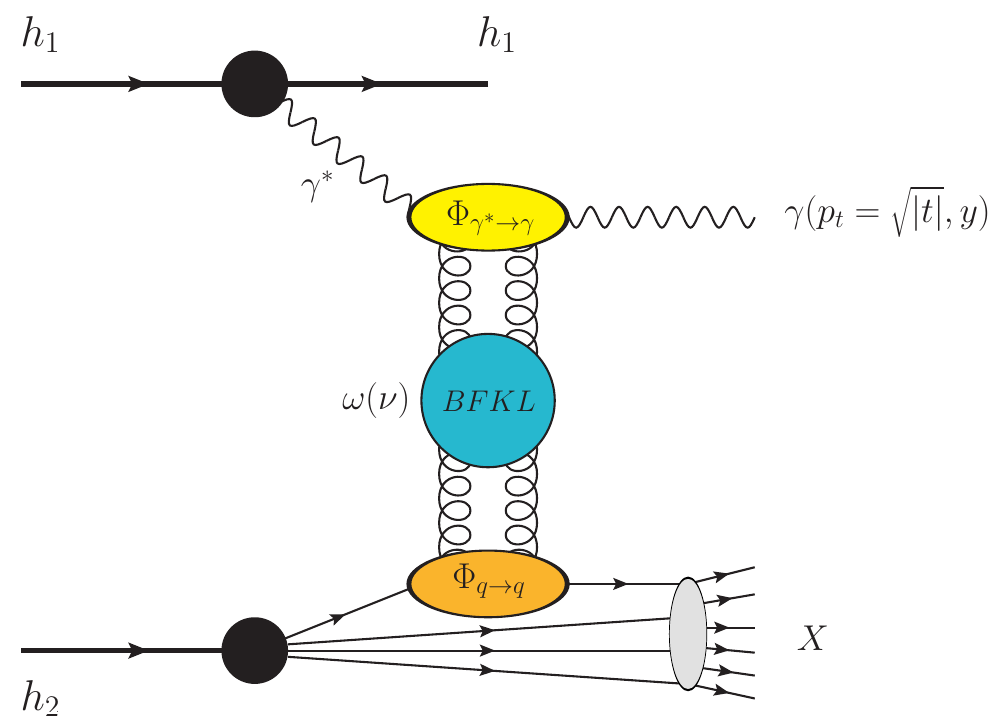}
\caption{Diffractive photon production at large momentum transfer in ultraperipheral collisions, $h_1 + h_2 \to h_1 \otimes \gamma \otimes X$. The photon emitted by $h_1$ couples to the BFKL ladder via the $\gamma^*\to\gamma$ impact factor, while the ladder couples to a parton from $h_2$ via the $q\to q$ impact factor. The struck parton (quark or gluon) initiates a jet that becomes part of the system $X$.}
\label{fig:diagram}
\end{figure}

In recent years, it has been established that photon-hadron interactions in hadronic collisions offer an effective approach for exploring QCD at high energies~\cite{Bertulani:2005ru,Klein:2020fmr}. In particular, findings regarding vector meson photoproduction at large momentum transfer indicate that the cross-sections for these processes in $pp$, $pA$, and $AA$ collisions are sufficiently large, making experimental analysis at the LHC, in principle, feasible~\cite{Frankfurt:2006tp, Frankfurt:2008er,Goncalves:2009gs,Goncalves:2010zb,Goncalves:2017sek, Bulhoes:2025pnp}. Motivated by these findings, this paper presents the first analysis of  diffractive  photoproduction of high $p_T$ photons  in ultraperipheral collisions (UPCs) at the LHC, which is represented in Fig.~\ref{fig:diagram}. The reaction $h_1 + h_2 \to h_1 \otimes \gamma \otimes X$, where $\otimes$ represents the presence of a rapidity gap in the final state,  can be factorized into the following stages: first, an incoming hadron $h_1$ emits a photon that fluctuates into a quark-antiquark ($q\bar{q}$) pair. Second, this pair interacts with a parton ($q$) from the target hadron via the exchange of a color-singlet state—represented by the BFKL ladder. Finally, the $q\bar{q}$ pair annihilates into a real photon with transverse momentum $p_T = \sqrt{|t|}$ and rapidity $y$. During the interaction, the ladder transfers momentum to the target parton, causing the hadron $h_2$ to dissociate and subsequently hadronize.  In the impact-factor representation, the scattering amplitude for $\gamma^* p \to \gamma X$ factorizes into the photon ($\Phi_{\gamma^* \to \gamma}$) and quark ($\Phi_{q \to q}$) impact factors and the Green function for two interacting reggeized gluons in a color-singlet configuration. At lowest order, the process is described by two-gluon exchange, yielding an energy-independent cross section. Beyond lowest order, the dominant contribution is associated with the BFKL Pomeron singularity generated by ladder diagrams with (reggeized) gluon exchange along the ladder. The exchange of an interacting gluon ladder then leads to a cross section that increases with energy, as determined by the characteristic function   $\omega(\nu)$ of the BFKL equation.

In this paper, we estimate diffractive photoproduction of high-$p_T$ photons in ultraperipheral $pp$, $p\mathrm{Pb}$, and $\mathrm{Pb}\mathrm{Pb}$ collisions at LHC energies, and we present predictions for the total cross section as well as for the rapidity and transverse-momentum distributions. We perform the calculations with different levels of sophistication in the treatment of the BFKL kernel. In particular, we investigate how the predictions depend on the prescription used to compute next-to-leading-order corrections to the characteristic function, thereby quantifying the associated theoretical uncertainty. We also confront our predictions with existing HERA data to constrain the only free parameter in the formalism—the BFKL energy scale $s_0$—for each kernel prescription. As a consequence, the resulting LHC predictions are  parameter free. The results presented below indicate a non-negligible yield of large-transverse-momentum photons produced in UPCs at the LHC, motivating future experimental searches for this final state as a probe of QCD dynamics at high energies.

This paper is organized as follows. In the next section, we present the formalism used to describe diffractive photoproduction of photons at large $t$ in UPCs. In particular, we discuss the scattering amplitude for $\gamma^* p \to \gamma X$ and summarize the different prescriptions for the BFKL kernel, as well as the ingredients entering our calculations. In Sec.~\ref{sec:res}, we use HERA data on the differential transverse-momentum distribution to constrain the BFKL energy scale $s_0$ and present the corresponding predictions for the energy dependence of the total cross section. We also present predictions for differential distributions and total cross sections considering  ultraperipheral $pp$, $p\mathrm{Pb}$, and $\mathrm{Pb}\mathrm{Pb}$ collisions at LHC. Finally, in Sec.~\ref{sec:conclusions}, we summarize our main results and conclusions.

%-----------------------------------------
\section{Formalism}
\label{sec:form}
%-----------------------------------------
In this section, we present the formalism used to describe diffractive photoproduction of photons in ultraperipheral collisions. In these events, the incident hadrons interact at impact parameters larger than the sum of their radii, so that strong interactions between them are strongly suppressed. Consequently, photon--photon and photon--hadron interactions dominate, and the hadronic cross section can be computed using the equivalent photon approximation~\cite{Budnev:1975poe,Bertulani:2005ru,Baltz:2007kq}. For the process illustrated in Fig.~\ref{fig:diagram}, the reaction reads
\begin{equation}
\label{eq:UPC_process}
h_1 + h_2 \,\to\, h_1 \otimes \gamma \otimes X \;,
\end{equation}
 and the differential cross section can be factorized as follows
\begin{equation}
\label{eq:UPC_dsig}
\frac{\drv\sigma\left[h_1 + h_2 \to h_1 \otimes \gamma \otimes X\right](\sqrt{s_{NN}})}{\drv y\, \drv t} \,=\, \omega\, \frac{\drv N_{\gamma/h_1}(\omega)}{\drv\omega}\; \frac{\drv\sigma_{\gamma^* h_2 \to \gamma X}\left(W\right)}{\drv t} \,\,,
\end{equation}
where $\sqrt{s_{NN}}$ is the hadron--hadron center-of-mass energy per nucleon pair. The basic idea is that the hadron $h_1$ emits a photon with energy $\omega$, with flux $\drv N_{\gamma/h_1}/\drv\omega$. This photon then interacts with the target hadron $h_2$ at a photon--hadron center-of-mass energy $W = \sqrt{2\omega\sqrt{s_{NN}}}$. The corresponding differential cross section is $\drv\sigma(\gamma^* h_2 \to \gamma X)/\drv t$. The rapidity of the final-state photon is $y \,\simeq\, \ln\!\left({2\omega}/{\sqrt{|t|}}\right)$, implying that forward rapidities probe large values of $W$. The main ingredients used in our calculations of Eq.~(\ref{eq:UPC_dsig}) will be presented in the following subsections.

\subsection{Equivalent Photon Flux}

Following the studies performed in Refs.    , in what follows we assume that  the photon spectrum for the proton is described by the expression derived in Ref.~\cite{Drees:1988pp}, which is the following
\begin{equation}
\label{eq:flux_p}
\begin{split}
\frac{\drv N_{\gamma/p}(\omega)}{\drv\omega} &\,=\, \frac{\alpha}{2\pi\omega} \left[ 1 + \left(1 - \frac{2\omega}{\sqrt{s}}\right)^2 \right] \left( \ln\Omega - \frac{11}{6} + \frac{3}{\Omega} - \frac{3}{2\Omega^2} + \frac{1}{3\Omega^3} \right) , \\[0.15cm]
\mbox{with } \hspace{1cm}
\Omega &\,=\, 1 + \frac{0.71~\text{GeV}^2}{Q^2_{\min}} \;, \hspace{1cm} \mbox{and} \qquad Q^2_{\min} \,=\, \frac{\omega^2}{\gamma_L^2\,(1 - 2\omega/\sqrt{s})} \;,
\end{split}
\end{equation}
where $\gamma_L = \sqrt{s_{NN}}/(2m_N)$ is the associated Lorentz factor and $m_N$ the nucleon mass. On the other hand, for the photon spectrum associated with a nucleus, we assume~\cite{Bertulani:2005ru}
\begin{equation}
\label{eq:flux_A}
\frac{\drv N_{\gamma/A}(\omega)}{\drv\omega} \,=\, \frac{2Z^2\alpha}{\pi\omega} \left[ \bar{\eta}\, K_0(\bar{\eta}) K_1(\bar{\eta}) - \frac{\bar{\eta}^2}{2}\, {\cal U}(\bar{\eta}) \right] ,
\end{equation}
where $\bar{\eta} = \omega\,(R_{h_1} + R_{h_2})/\gamma_L$, ${\cal U}(\bar{\eta}) = K_1^2(\bar{\eta}) - K_0^2(\bar{\eta})$, and $K_{0,1}$ are modified Bessel functions, with $R_A = 1.2\,A^{1/3}$~fm and $R_p = 0.7$~fm.
No additional gap-survival factor is included. The suppression of hadronic interactions enters only through the condition $b > R_{h_1} + R_{h_2}$ in the nuclear flux.

In our calculations, we assume that hadron $h_1$ moves toward positive rapidities. For symmetric systems (\emph{i.e.} $pp$ and PbPb), we include both photon-emission possibilities by adding the contribution with $y\to -y$, which yields a symmetric rapidity distribution. For $p$Pb collisions, we assume that the intact hadron can be tagged and the photon emitter can therefore be identified. In what follows, we denote by $\mathrm{Pb}p$ the configuration in which the nucleus emits the photon and the proton dissociates, and by $p\mathrm{Pb}$ the one in which the proton emits the photon and the nucleus dissociates.

%-----------------------------------------
\subsection{The $\gamma^* h \to \gamma X$ process}
\label{ssec:process}
%-----------------------------------------

At large momentum transfer, the photon-hadron process $\gamma^* h \to \gamma X$ with a rapidity gap between the outgoing photon and the system $X$ is described as the elastic scattering of the photon on a single parton of the hadron, $\gamma^* q \to \gamma q$ (or $\gamma^* g \to \gamma g$), through the exchange of a color-singlet BFKL ladder in the $t$-channel~\cite{Mueller:1992pe,Evanson:1999zb}.
The hard scale is provided by the momentum transfer, $|t| = p_T^2$, with $p_T$ the transverse momentum of the photon.
In the high-energy regime, the $\gamma^* h \to \gamma X$ cross section at photon-hadron center-of-mass energy $W$ is given by the convolution of the partonic cross section, $ {\drv\hat{\sigma}}/{\drv t}(\gamma^* q \to \gamma q)$, with the parton distributions of the hadron $h$, as follows~\cite{Mueller:1992pe,Goncalves:2013pac}
\begin{equation}
\label{eq:feff}
\frac{\drv\sigma(\gamma^* h \to \gamma X)}{\drv t\, \drv x_j} \,=\, \left[ \frac{81}{16}\, G_h(x_j,|t|) + \sum_j \left( q_h^j(x_j,|t|) + \bar{q}_h^j(x_j,|t|) \right) \right] \frac{\drv\hat{\sigma}(\gamma^* q \to \gamma q)}{\drv t}(\shat) \;,
\end{equation}
where $G_h$ and $q_h^j$ are the gluon and quark distributions, evaluated at the factorization scale $|t|$, and $x_j$ is the fraction of the longitudinal momentum of the hadron carried by the struck parton, which initiates a jet, so that the squared photon-parton center-of-mass energy is given by $\shat = x_j W^2$.
The single differential cross section, ${\drv\sigma}/{\drv t}$, needed to calculate Eq.~(\ref{eq:UPC_dsig}), is given by
\begin{equation}
\label{eq:dsig_gp}
\frac{\drv\sigma}{\drv t} \,=\, \int_{x_{\min}}^{1} \drv x_j\; \frac{\drv\sigma(\gamma^* h \to \gamma X)}{\drv t\, \drv x_j} \;,
\end{equation}
with the minimum value $x_{\min}$ being determined by the experimental cut on the mass $M_X$ of the hadron remnants. Following the selection used in the 
H1 measurement~\cite{H1:2008jdn}, we impose $\Delta\eta \gtrsim 3$, where $\Delta\eta$ is the rapidity
gap between the photon and the system  $X$, which corresponds to $\yIP=|t|/\hat{s}<0.05$ and implies  $ x_{\min} \,=\, {|t|}/({0.05\,W^2})$

In our calculations for a proton target we employ the {\tt MMHT2014nlo68cl} proton PDFs~\cite{Harland-Lang:2014zoa}, accessed through {\tt LHAPDF}~\cite{Buckley:2014ana}. On the other hand, for the dissociating nucleus, the parton distributions in Eq.~\eqref{eq:feff} are $A$ times the nuclear PDFs per nucleon of {\tt EPPS21} parameterization~\cite{Eskola:2021nhw}. The collision systems, with the center-of-mass energies, photon emitters, PDFs and radii used in this work, are summarized in Table~\ref{tab:systems}.

\begin{table*}[!t]
\centering
\setlength{\tabcolsep}{10pt}
\renewcommand{\arraystretch}{1.30}
\begin{tabular}{lcccc}
\toprule
\textbf{System} & $\sqrt{s}$ [TeV] & photon emitter & dissociating hadron (PDF) & $R_{h_1}+R_{h_2}$ [fm] \\
\midrule
$pp$  & 13.6 & both & $p$ ({\tt MMHT2014}) & --- \\
Pb$p$ & 8.16 & Pb   & $p$ ({\tt MMHT2014}) & 7.81 \\
$p$Pb & 8.16 & $p$  & Pb ($A\times${\tt EPPS21}) & --- \\
PbPb  & 5.36 & both & Pb ($A\times${\tt EPPS21}) & 14.22 \\
\bottomrule
\end{tabular}
\caption{Collision systems considered.
In the asymmetric configurations, the photon emitter is the tagged intact hadron.
The last column gives the minimum impact parameter entering the nuclear photon flux, Eq.~\eqref{eq:flux_A}, with $R_p = 0.7$~fm and $R_A = 1.2\,A^{1/3}$~fm; it is not used when the photon is emitted by the proton, whose flux, Eq.~\eqref{eq:flux_p}, does not depend on the impact parameter.}
\label{tab:systems}
\end{table*}

%In Ref.~\cite{H1:2008jdn}, $\yIP = p\cdot(q - p_\gamma)/(p\cdot q)$ is defined as the inelasticity of the $\gamma^* p$ interaction, with $p$, $q$ and $p_\gamma$ the four-momenta of the proton, of the incoming and of the outgoing photon, and $\yIP \simeq e^{-\Delta\eta}$, with $\Delta\eta$ the rapidity gap between the photon and the system $X$, so that $\yIP < 0.05$ corresponds to $\Delta\eta \gtrsim 3$.For the partonic subprocess, with the struck parton carrying the momentum $x_j\,p$, it reduces to $\yIP = |t|/(x_j W^2) = |t|/\shat$. In Ref.~\cite{Goncalves:2013pac}, a fixed value $x_{\min} = 0.01$ was used instead. We employ the {\tt MMHT2014nlo68cl} proton PDFs~\cite{Harland-Lang:2014zoa}, accessed through {\tt LHAPDF}~\cite{Buckley:2014ana}.

%-----------------------------------------
\subsection{Photon production at large $|t|$ in the BFKL approach}
\label{ssec:amplitude}
%-----------------------------------------

Following Refs.~\cite{Evanson:1999zb,Goncalves:2013pac}, the cross section of the partonic subprocess $\gamma^* q \to \gamma q$ is
\begin{equation}
\label{eq:dsighat}
\frac{\drv\hat{\sigma}}{\drv t}(\gamma^* q \to \gamma q) \,=\, \frac{1}{16\pi\shat^2} \left\{ |{\cal A}_{(+,+)}(\shat,t)|^2 + |{\cal A}_{(+,-)}(\shat,t)|^2 \right\} ,
\end{equation}
where $\shat$ is the center-of-mass energy squared of the subprocess.
For an incoming photon of virtuality $Q^2$ and the lowest conformal spin, the helicity amplitudes read~\cite{Evanson:1999zb,Goncalves:2013pac}
\begin{equation}
\label{eq:amplitudes_Q2}
\begin{split}
{\cal A}_{(+,+)}(\shat,t) \,=\, & \; i\alpha\as^2 \sum_q e_q^2\, \frac{\pi}{6}\, \frac{\shat}{|t|} \int \drv\nu\, \frac{\nu^2}{(\nicefrac{1}{4}+\nu^2)^2}\, \frac{\tanh(\pi\nu)}{\pi\nu(1+\nu^2)} \left[\frac{\shat}{s_0}\right]^{\omega(\nu)} \int_{\nicefrac{1}{2}-i\infty}^{\nicefrac{1}{2}+i\infty} \frac{\drv z}{2\pi i} \left[\frac{Q^2}{|t|}\right]^{-z/2} \\[0.10cm]
& \times \frac{\Gamma(\nicefrac{1}{2}-i\nu-\nicefrac{z}{2})\,\Gamma(\nicefrac{1}{2}+i\nu-\nicefrac{z}{2})}{|\Gamma(\nicefrac{1}{2}+i\nu)|^2}\, \Gamma(\nicefrac{z}{2})\,\Gamma(\nicefrac{z}{2}+1) \left[ z^2 + 11 + 12\nu^2 \right] , \\[0.20cm]
{\cal A}_{(+,-)}(\shat,t) \,=\, & \; i\alpha\as^2 \sum_q e_q^2\, \frac{2\pi}{3}\, \frac{\shat}{|t|} \int \drv\nu\, \frac{\nu^2}{(\nicefrac{1}{4}+\nu^2)^2}\, \frac{\tanh(\pi\nu)}{\pi\nu(1+\nu^2)} \left[\frac{\shat}{s_0}\right]^{\omega(\nu)} \int_{\nicefrac{1}{2}-i\infty}^{\nicefrac{1}{2}+i\infty} \frac{\drv z}{2\pi i} \left[\frac{Q^2}{|t|}\right]^{-z/2} \\[0.10cm]
& \times \frac{\Gamma(\nicefrac{3}{2}-i\nu-\nicefrac{z}{2})\,\Gamma(\nicefrac{3}{2}+i\nu-\nicefrac{z}{2})}{|\Gamma(\nicefrac{1}{2}+i\nu)|^2}\, \Gamma(\nicefrac{z}{2})\,\Gamma(\nicefrac{z}{2}+1) \;.
\end{split}
\end{equation}
In the photoproduction regime, $Q^2 \ll |t|$, which is the regime of our interest, the $z$ integrals are dominated by the pole of $\Gamma(z/2)$ at $z = 0$, and the amplitudes become~\cite{Ivanov:1998jw,Goncalves:2013pac}
\begin{equation}
\label{eq:amplitudes}
\begin{split}
{\cal A}_{(+,+)}(\shat,t) &\,=\, i\alpha\as^2 \sum_q e_q^2\, \frac{4\pi}{3}\, \frac{\shat}{|t|} \int \drv\nu\, \frac{\nu^2}{(1/4+\nu^2)^2}\, \frac{\tanh(\pi\nu)}{\pi\nu}\, \frac{11/4+3\nu^2}{1+\nu^2} \left[\frac{\shat}{s_0}\right]^{\omega(\nu)} , \\[0.15cm]
{\cal A}_{(+,-)}(\shat,t) &\,=\, i\alpha\as^2 \sum_q e_q^2\, \frac{4\pi}{3}\, \frac{\shat}{|t|} \int \drv\nu\, \frac{\nu^2}{(1/4+\nu^2)^2}\, \frac{\tanh(\pi\nu)}{\pi\nu}\, \frac{1/4+\nu^2}{1+\nu^2} \left[\frac{\shat}{s_0}\right]^{\omega(\nu)} ,
\end{split}
\end{equation}
with $\alpha$ the electromagnetic coupling, $e_q$ the quark charges, $s_0$ the BFKL energy scale and $\omega(\nu)$ the BFKL characteristic function, which will be discussed in Sec.~\ref{ssec:kernels}.
The two-gluon exchange (Born) limit corresponds to $\omega(\nu)=0$.

The amplitudes in Eq.~\eqref{eq:amplitudes} contain the impact factors at leading order (LO).
As in Ref.~\cite{Goncalves:2013pac}, the strong coupling entering the impact factors is kept fixed, $\as = 0.21$, for all the kernels considered below, while it runs in the characteristic function for the NLA approaches.
The BFKL energy scale is parametrized as
\begin{equation}
\label{eq:s0}
s_0 \,=\, \cso\, |t| \;,
\end{equation}
where $\cso$ is a free parameter,  which will be determined for each kernel using H1 data . This corresponds to the parametrization $s_0 = c_1 Q^2 + c_2 |t|$ of Ref.~\cite{Goncalves:2013pac} in the limit of quasi-real photons, $Q^2 \simeq 0$. In our analysis, we consider four active flavors in the photon loop, which implies $\sum_q e_q^2 = 10/9$.

%-----------------------------------------
\subsection{The BFKL characteristic function}
\label{ssec:kernels}
%-----------------------------------------

The BFKL characteristic function is written as~\cite{Goncalves:2013pac}
\begin{equation}
\label{eq:omega_LLA}
\omega(\nu) \,=\, \abar\, \chi(\gamma) \;, \qquad \gamma \,=\, \tfrac{1}{2} + i\nu \;,
\end{equation}
with $\abar \equiv N_c\as/\pi$.
In the leading logarithmic approximation (LLA)~\cite{Fadin:1975cb,Kuraev:1976ge,Kuraev:1977fs,Balitsky:1978ic},
\begin{equation}
\label{eq:chiLO}
\chi^{\LO}(\gamma) \,=\, 2\psi(1) - \psi(\gamma) - \psi(1-\gamma) \;,
\end{equation}
with $\psi$ the digamma function.

In the next-to-leading logarithmic approximation (NLA)~\cite{Fadin:1998py,Ciafaloni:1998gs},
\begin{equation}
\label{eq:omega_NLA}
\chi(\gamma) \,=\, \chi^{\LO}(\gamma) + \abar\, \chi^{\NLO}(\gamma) \;,
\end{equation}
where
\begin{equation}
\label{eq:chiNLO}
\begin{split}
\chi^{\NLO}(\gamma) \,=\, & \;\frac{4 - \pi^2 + 5\beta_0/N_c}{12}\,\chi^{\LO}(\gamma) + \frac{1}{4}\left[\psi''(\gamma) + \psi''(1-\gamma)\right] - \frac{1}{4}\left[\phi(\gamma) + \phi(1-\gamma)\right] \\[0.10cm]
& - \frac{\pi^2 \cos(\pi\gamma)}{4\sin^2(\pi\gamma)(1-2\gamma)} \left\{ 3 + \left(1+\frac{N_f}{N_c^3}\right) \frac{2+3\gamma(1-\gamma)}{(3-2\gamma)(1+2\gamma)} \right\} \\[0.10cm]
& + \frac{3}{2}\zeta(3) - \frac{\beta_0}{8N_c}\left(\chi^{\LO}(\gamma)\right)^2 ,
\end{split}
\end{equation}
with $\beta_0 = (11N_c - 2N_f)/3$ and
\begin{equation}
\label{eq:phi}
\phi(\gamma) + \phi(1-\gamma) \,=\, \sum_{m=0}^{\infty} \left[ \frac{1}{\gamma+m} + \frac{1}{1-\gamma+m} \right] \left[ \psi'\!\left(\frac{2+m}{2}\right) - \psi'\!\left(\frac{1+m}{2}\right) \right] .
\end{equation}
Eq.~\eqref{eq:chiNLO} corresponds to the $\MSb$ scheme with renormalization scale $\muR^2 = |t|$.
For a generic scale, the NLA characteristic function becomes~\cite{Ivanov:2014hpa}
\begin{equation}
\label{eq:omega_NLA_mu}
\omega(\nu,\muR) \,=\, \abar(\muR)\, \chi^{\LO}(\gamma) + \abar^2(\muR) \left[ \chi^{\NLO}(\gamma) + \frac{\beta_0}{4N_c}\, \chi^{\LO}(\gamma) \ln\frac{\muR^2}{|t|} \right] .
\end{equation}
The coupling is evaluated at one loop,
\begin{equation}
\label{eq:alphas}
\as(\mu^2) \,=\, \frac{4\pi}{\beta_0 \ln(\mu^2/\LQCD^2)} \;,
\end{equation}
with $N_f = 4$ and $\LQCD = 0.121$~GeV, which corresponds to $\as(M_Z) = 0.118$ matched at $m_b = 4.8$~GeV.
As shown in Refs.~\cite{Fadin:1998py,Ciafaloni:1998gs}, the NLA correction $\chi^{\NLO}$ is large and negative at small $\nu$, so that the pure NLA characteristic function becomes negative around $\nu = 0$ (see Fig.~\ref{fig:omega}), and the perturbative series must be stabilized.
In what follows, we consider three alternative prescriptions, which differ only by terms beyond the NLA accuracy.

\subsubsection{All-poles collinear resummation}

In the All-poles approach~\cite{SabioVera:2005tiv}, the collinear poles of the NLA kernel at $\gamma = 0$ and 1 are resummed to all orders, and the characteristic function is~\cite{Goncalves:2013pac}
\begin{equation}
\label{eq:omega_AP}
\begin{split}
\omega_{\rm All-poles} \,=\, & \;\abar\,\chi^{\LO}(\gamma) + \abar^2\chi^{\NLO}(\gamma)
+ \Bigg\{ \sum_{m=0}^{\infty} \Bigg[ \left( \sum_{n=0}^{\infty} \frac{(-1)^n (2n)!}{2^n n! (n+1)!} \frac{(\abar + a\abar^2)^{n+1}}{(\gamma + m - b\abar)^{2n+1}} \right) \\
&  - \frac{\abar}{\gamma+m}- \abar^2 \left( \frac{a}{\gamma+m} + \frac{b}{(\gamma+m)^2} - \frac{1}{2(\gamma+m)^3} \right) \Bigg] + \{\gamma \to 1-\gamma\} \Bigg\} ,
\end{split}
\end{equation}
where
\begin{equation}
\label{eq:ab_AP}
a \,=\, \frac{5\beta_0}{12N_c} - \frac{13 N_f}{36 N_c^3} - \frac{55}{36} \;, \qquad
b \,=\, -\frac{\beta_0}{8N_c} - \frac{N_f}{6N_c^3} - \frac{11}{12} \;,
\end{equation}
and $\abar = \abar(|t|)$.
The sum over $n$ is evaluated in a closed form, given by
\begin{equation}
\label{eq:AP_closed}
\sum_{n=0}^{\infty} \frac{(-1)^n (2n)!}{2^n n! (n+1)!} \frac{(\abar + a\abar^2)^{n+1}}{(\gamma + m - b\abar)^{2n+1}} \,=\, \sqrt{(\gamma+m-b\abar)^2 + 2(\abar+a\abar^2)} - (\gamma+m-b\abar) \;.
\end{equation}

\subsubsection{BLM scale setting}

In the Brodsky-Lepage-Mackenzie (BLM) approach~\cite{Brodsky:1982gc}, the renormalization scale is chosen so as to absorb the $\beta_0$-dependent terms into the running coupling.
Following Ref.~\cite{Brodsky:1998kn}, this is done in a physical (MOM) scheme, and the result is then expressed in terms of the $\MSb$ coupling~\cite{Ivanov:2014hpa}.
The procedure consists of four steps.

\textit{(i) Conformal and $\beta_0$-dependent parts.}
The NLA correction of Eq.~\eqref{eq:chiNLO} is separated as~\cite{Brodsky:1998kn}
\begin{equation}
\label{eq:chiNLO_split}
\chi^{\NLO}(\gamma) \,=\, \bar{\chi}(\gamma) + \frac{\beta_0}{4N_c}\, \chi^{\LO}(\gamma) \left( \frac{5}{3} - \frac{\chi^{\LO}(\gamma)}{2} \right) ,
\end{equation}
where, from Eq.~\eqref{eq:chiNLO}, the conformal part, which does not depend on $\beta_0$, is
\begin{equation}
\label{eq:chibar}
\begin{split}
\bar{\chi}(\gamma) \,=\, & \;\frac{4 - \pi^2}{12}\,\chi^{\LO}(\gamma) + \frac{1}{4}\left[\psi''(\gamma) + \psi''(1-\gamma)\right] - \frac{1}{4}\left[\phi(\gamma) + \phi(1-\gamma)\right] \\[0.10cm]
& - \frac{\pi^2 \cos(\pi\gamma)}{4\sin^2(\pi\gamma)(1-2\gamma)} \left\{ 3 + \left(1+\frac{N_f}{N_c^3}\right) \frac{2+3\gamma(1-\gamma)}{(3-2\gamma)(1+2\gamma)} \right\} + \frac{3}{2}\zeta(3) \;.
\end{split}
\end{equation}
Inserting Eq.~\eqref{eq:chiNLO_split} in Eq.~\eqref{eq:omega_NLA_mu}, the $\MSb$ characteristic function reads
\begin{equation}
\label{eq:omega_MS}
\omega(\nu,\muR) \,=\, \abar^{\MSb}\, \chi^{\LO} + \big(\abar^{\MSb}\big)^2 \left[ \bar{\chi} + \frac{\beta_0}{4N_c}\, \chi^{\LO} \left( \ln\frac{\muR^2}{|t|} + \frac{5}{3} - \frac{\chi^{\LO}}{2} \right) \right] ,
\qquad \abar^{\MSb} \equiv \abar^{\MSb}(\muR) \;.
\end{equation}

\textit{(ii) Change of scheme.}
The transition to the MOM scheme is a finite renormalization of the coupling~\cite{Brodsky:1998kn,Celmaster:1979km},
\begin{equation}
\label{eq:MOM}
\begin{gathered}
\as^{\MSb} \,=\, \as^{\rm MOM} \left[ 1 + T\, \frac{\as^{\rm MOM}}{\pi} \right] , \qquad
T \,=\, T^{\rm conf} + T^\beta \;, \\[0.15cm]
T^{\rm conf} \,=\, \frac{N_c}{8}\,\frac{17}{2}\, I \;, \qquad
T^\beta \,=\, -\frac{\beta_0}{2} \left( 1 + \frac{2}{3} I \right) ,
\end{gathered}
\end{equation}
for the MOM scheme in the Landau gauge, with $I = -2\int_0^1 \drv x\, \ln x/(x^2-x+1) \simeq 2.3439$.
Since $\as/\pi = \abar/N_c$, inserting Eq.~\eqref{eq:MOM} in Eq.~\eqref{eq:omega_MS} gives, at NLA accuracy,
\begin{equation}
\label{eq:omega_MOM}
\begin{split}
\omega(\nu,\muR) \,=\, & \abar^{\rm MOM}\, \chi^{\LO} + \big(\abar^{\rm MOM}\big)^2 \bigg[ \bar{\chi} + \frac{T^{\rm conf}}{N_c}\, \chi^{\LO} \\
& + \frac{\beta_0}{4N_c}\, \chi^{\LO} \left( \ln\frac{\muR^2}{|t|} - 2\left(1+\frac{2}{3}I\right) + \frac{5}{3} - \frac{\chi^{\LO}}{2} \right) \bigg] .
\end{split}
\end{equation}

\textit{(iii) BLM scale.}
The $\beta_0$-dependent term of Eq.~\eqref{eq:omega_MOM} vanishes for
\begin{equation}
\label{eq:mu_BLM_nu}
\hat{Q}^2(\nu) \,=\, |t| \exp\left[ 2\left(1+\frac{2}{3}I\right) - \frac{5}{3} + \frac{\chi^{\LO}(\gamma)}{2} \right] ,
\end{equation}
which is the $\nu$-dependent scale of Refs.~\cite{Brodsky:1998kn,Goncalves:2013pac}.
In order to avoid the Landau pole of this scale at large $\nu$, we adopt the $\nu$-independent scale (variant $a$ of Ref.~\cite{Ivanov:2014hpa}), which absorbs only the $\nu$-independent terms,
\begin{equation}
\label{eq:mu_BLM}
\mu_{\BLM}^2 \,=\, |t| \exp\left[ 2\left(1+\frac{2}{3}I\right) - \frac{5}{3} \right] \simeq 32\, |t| \;,
\end{equation}
so that
\begin{equation}
\label{eq:omega_BLM_MOM}
\omega_{\BLM}(\nu) \,=\, \abar^{\rm MOM}(\mu_{\BLM})\, \chi^{\LO} + \big(\abar^{\rm MOM}(\mu_{\BLM})\big)^2 \left[ \bar{\chi} + \frac{T^{\rm conf}}{N_c}\, \chi^{\LO} - \frac{\beta_0}{8N_c} \left(\chi^{\LO}\right)^2 \right] .
\end{equation}

\textit{(iv) Return to the $\MSb$ scheme.}
Inverting Eq.~\eqref{eq:MOM}, $\as^{\rm MOM} = \as^{\MSb}\,[1 - T\,\as^{\MSb}/\pi]$, at the same scale $\mu_{\BLM}$, the term $T^{\rm conf}$ cancels and one obtains~\cite{Ivanov:2014hpa}
\begin{equation}
\label{eq:omega_BLM}
\omega_{\BLM}(\nu) \,=\, \abar^{\MSb}(\mu_{\BLM})\, \chi^{\LO} + \big(\abar^{\MSb}(\mu_{\BLM})\big)^2 \left[ \bar{\chi} - \frac{T^\beta}{N_c}\, \chi^{\LO} - \frac{\beta_0}{8N_c} \left(\chi^{\LO}\right)^2 \right] ,
\end{equation}
which is the expression used in this work, with the one-loop $\MSb$ coupling of Eq.~\eqref{eq:alphas}.

\begin{table}[t]
\centering
\footnotesize
\setlength{\tabcolsep}{8pt}
\renewcommand{\arraystretch}{1.30}
\begin{tabular}{c|ccccccc}
\toprule
 & \multicolumn{7}{c}{$Y$} \\
\cmidrule(l){2-8}
$|t|$ [GeV$^2$] & 1 & 2 & 3 & 5 & 7 & 10 & 12 \\
\midrule
1   & 10.8 & 54.6 & 90.0 & 148 & 245 & 259 & 252 \\
4   & 6.4 & 3.3 & 4.9 & 8.1 & 11.4 & 16.6 & 20.3 \\
6   & 9.0 & 2.2 & 3.2 & 5.1 & 7.1 & 10.2 & 12.4 \\
12  & 13.8 & 4.0 & 2.5 & 2.5 & 3.5 & 4.9 & 5.9 \\
25  & 20.1 & 5.9 & 4.8 & 5.1 & 6.4 & 9.2 & 11.4 \\
36  & 23.8 & 6.7 & 5.5 & 6.2 & 7.9 & 11.6 & 14.5 \\
100 & 37.5 & 9.0 & 7.3 & 8.3 & 10.8 & 16.1 & 20.5 \\
\bottomrule
\end{tabular}
\caption{PMS scale $\muR^2/|t|$ as a function of $|t|$ and $Y = \ln(\shat/s_0)$.}
\label{tab:PMS}
\end{table}

\subsubsection{Principle of minimal sensitivity}

According to the principle of minimal sensitivity (PMS) \cite{Stevenson:1980du, Stevenson:1981vj}, the renormalization scale is selected at the value for which the observable exhibits the least sensitivity to its variation. 
As in Refs.~\cite{Ivanov:2005gn,Caporale:2008is,Ivanov:2014hpa}, the optimization is performed on the observable itself, here the partonic cross section~\eqref{eq:dsighat}, computed with the NLA characteristic function~\eqref{eq:omega_NLA_mu}.
Defining $L \equiv \ln(\muR^2/|t|)$ and $Y \equiv \ln(\shat/s_0)$, the optimal scale $L^*(|t|,Y)$ satisfies
\begin{equation}
\label{eq:PMS}
\frac{\partial}{\partial L} \ln \frac{\drv\hat{\sigma}}{\drv t}(Y,|t|;L) \bigg|_{L = L^*} \,=\, 0 \;.
\end{equation}
Since the impact factors are taken at LO with fixed coupling, the $\muR$ dependence comes entirely from the kernel.
Using $\drv\abar/\drv L = -(\beta_0/4N_c)\,\abar^2$, one finds
\begin{equation}
\label{eq:domega_dL}
\frac{\partial \omega}{\partial L} \,=\, -\frac{\beta_0}{2N_c}\, \abar^3 \left[ \chi^{\NLO}(\gamma) + \frac{\beta_0}{4N_c}\, \chi^{\LO}(\gamma)\, L \right] ,
\end{equation}
so that the optimal scale depends on both $|t|$ and $Y$.
Contrary to Refs.~\cite{Caporale:2008is,Ivanov:2014hpa}, we do not optimize $s_0$.
With LO impact factors, the cross section decreases monotonically with $s_0$, through $(\shat/s_0)^{\omega(\nu)}$, and has no stationary point.
Therefore, $s_0 = \cso|t|$ is fitted to the H1 data, as for the other kernels.

For each value of $|t|$, the optimal scale is determined on a grid in $Y$ with spacing $\Delta Y = 0.25$, and linearly interpolated between the grid points.
When the cross section has more than one stationary point, we choose the smoothest one, as in Ref.~\cite{Caporale:2008is}.
When no stationary point exists, we take the point of minimal sensitivity, where $|\partial \ln(\drv\hat{\sigma}/\drv t)/\partial L|$ is smallest.
The resulting scales are given in Table~\ref{tab:PMS}, with typical values $\muR^2 \simeq (2\text{--}20)\,|t|$ increasing with $Y$.

\subsubsection{Comparison between characteristic functions}
\begin{figure*}[!t]
\centering
\includegraphics[width=0.60\textwidth]{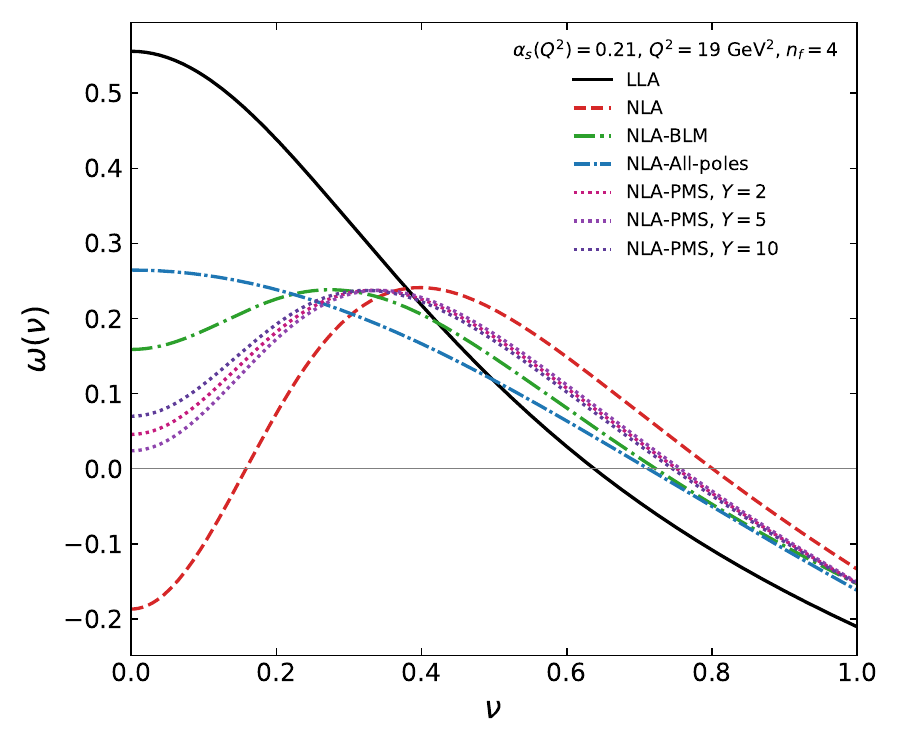}
\caption{Characteristic function $\omega(\nu)$ in the LLA, Eqs.~\eqref{eq:omega_LLA}--\eqref{eq:chiLO}, and in the NLA with $\muR^2 = Q^2$, Eq.~\eqref{eq:omega_NLA_mu}, together with the BLM~\eqref{eq:omega_BLM}, All-poles~\eqref{eq:omega_AP} and PMS prescriptions.
The scale is $Q^2 \simeq 19$~GeV$^2$, for which $\as(Q^2) = 0.21$.
The PMS eigenvalue is shown for three values of $Y = \ln(\shat/s_0)$, since the optimal scale depends on the energy.}
\label{fig:omega}
\end{figure*}

The characteristic functions are compared in Fig.~\ref{fig:omega} for $|t| \simeq 19$~GeV$^2$, where $\as = 0.21$.
The pure NLA function is negative at small $\nu$, while the three prescriptions give positive values, smaller than the LLA ones.
Since the region of small $\nu$ dominates at large $Y$, the value of $\omega(0)$ controls the growth of the cross section with energy, which is largest for the All-poles and smallest for the PMS prescription. In the next section, such prescriptions for the characteristic function will be used to estimate the diffractive photoproduction of photons with large transverse momentum in UPCs.

\FloatBarrier

%-----------------------------------------
\section{Results}
\label{sec:res}
%-----------------------------------------

\begin{table*}[t]
	\centering
	\setlength{\tabcolsep}{10pt}
	\renewcommand{\arraystretch}{1.30}
	\begin{tabular}{lcc}
		\toprule
		\textbf{Kernel} & $\cso$ & $\chi^2_{\drv\sigma/\drv|t|}$ \\
		\midrule
		LLA          & 77.4 & 6.82  \\
		NLA-BLM      & 56.1 & 4.97 \\
		NLA-All-poles& 38.3 & 4.12 \\
		NLA-PMS      & 80.4 & 4.22  \\
		\bottomrule
	\end{tabular}
	\caption{Results of the fit of $s_0 = \cso|t|$ to the H1 $\drv\sigma/\drv|t|$ data at $W = 219$~GeV~\cite{H1:2008jdn}.}
	\label{tab:fit}
\end{table*}

\begin{figure*}[h!]
	\centering
	\makebox[\textwidth][c]{%
		\includegraphics[width=0.51\textwidth]{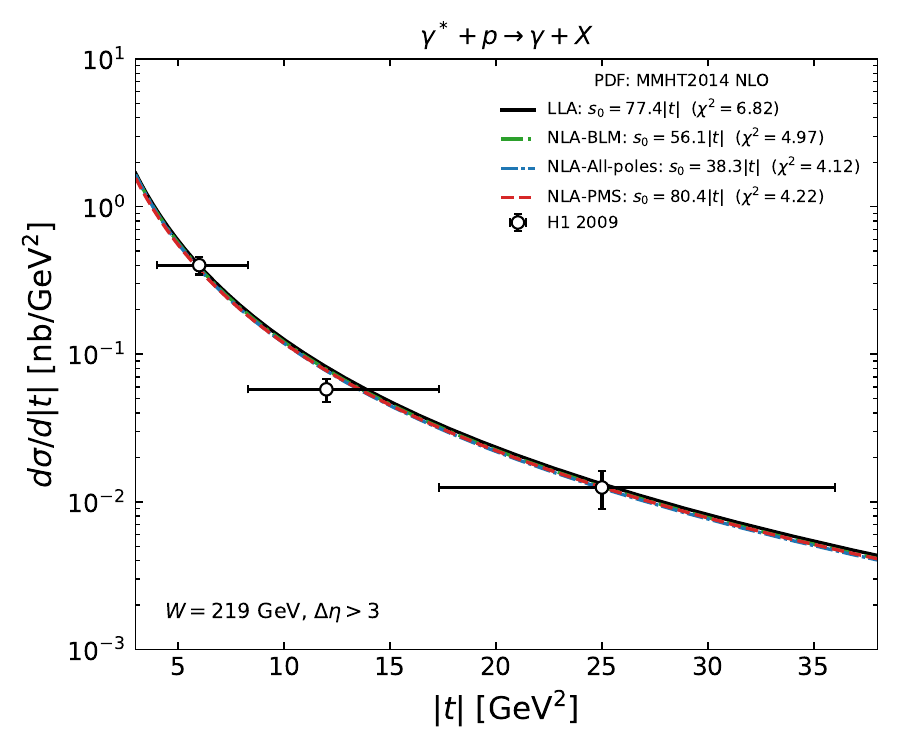}%
		\hspace{-0.15cm}%
		\includegraphics[width=0.51\textwidth]{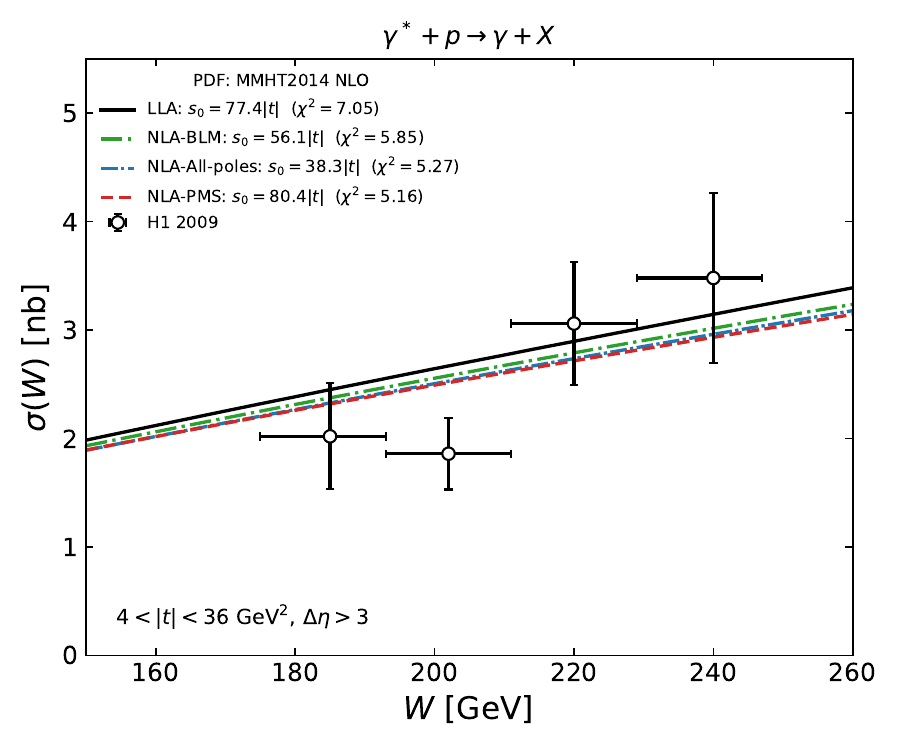}}
	\caption{Comparison with the H1 data~\cite{H1:2008jdn} for $\gamma^* p \to \gamma X$.
		{\bf Left:} $\drv\sigma/\drv|t|$ at $W = 219$~GeV, used to fix $s_0 = \cso|t|$ for each kernel.
		{\bf Right:} Predictions for the energy dependence of the integrated cross section $\sigma(W)$ over the range $4 < |t| < 36$~GeV$^2$.
		The values of $\cso$ and the $\chi^2$ for each curve are given in the legends.}
	\label{fig:H1}
\end{figure*}

Initially, to constrain the BFKL scale $s_0$, we compare our predictions with the experimental data measured in $ep$ collisions at HERA. In particular, the H1 collaboration has measured $\drv\sigma/\drv|t|$ at $W = 219$~GeV in three bins of $|t|$ between 4 and 36~GeV$^2$, and the cross section $\sigma(W)$ integrated over $4 < |t| < 36$~GeV$^2$ in four bins of $W$ between 175 and 247~GeV~\cite{H1:2008jdn}. As in Ref.~\cite{Goncalves:2013pac}, we  perform a $\chi^2$ minimization of our predictions for $\drv\sigma/\drv|t|$, whose results  
 are compared with the data as averages over the corresponding bins.
The parameter $\cso$ of each kernel is fitted to the three $\drv\sigma/\drv|t|$ points, by minimizing
\begin{equation}
	\label{eq:chi2}
	\chi^2 \,=\, \sum_{i=1}^{3} \left( \frac{\langle \drv\sigma/\drv|t| \rangle_i^{\rm th} - \langle \drv\sigma/\drv|t| \rangle_i^{\rm H1}}{\delta_i} \right)^2 ,
\end{equation}
where $\delta_i = \sqrt{(\delta_i^{\rm stat})^2 + (\delta_i^{\rm syst})^2}$ is the total experimental uncertainty of the $i$-th data point, while $\sigma(W)$ is a prediction.
The results of the fit are given in Table~\ref{tab:fit} and the corresponding distributions are compared with the H1 data in the left panel of Fig.~\ref{fig:H1}. The right panel compares our predictions for the integrated cross section with H1 data, obtained by integrating the fitted $d\sigma/d|t|$ over $4 < |t| < 36$ GeV$^2$. Predictions derived from different prescriptions for the BFKL characteristic function adequately describe the HERA data. However, limited data availability in $ep$ collisions precludes more definitive conclusions.

In what follows, we  present our predictions for UPCs at the LHC. Our work is primarily motivated by the significant increase in maximum photon-hadron center-of-mass energies in $pp$, $pPb$, and $PbPb$ collisions, which exceed 8400, 1400, and 950 GeV, respectively~\cite{Baltz:2007kq}. This represents an order of magnitude increase compared to the ~250 GeV range of the H1 measurement. Consequently, investigating diffractive photoproduction of high-$p_T$ photons in UPCs at the LHC allows us to probe the contribution of QCD dynamics for this process across an entirely new kinematic regime. 

Table \ref{tab:sigma} presents our integrated cross section predictions for various collision systems, utilizing distinct prescriptions for the BFKL characteristic function. These results were derived by integrating over the photon rapidity and the range $1 < |t| < 100$ GeV$^2$. Consistent with the H1 analysis, we impose a rapidity gap constraint of $\Delta\eta > 3.0$ between the photon and the system $X$. The $pp$ collision cross sections are on the order of $10$ nb and show sensitivity to the chosen BFKL characteristic function. As expected from previous studies, the Born term gives the lower bound for the cross section, and the LLA kernel the largest prediction, except for PbPb collisions, where the All-poles kernel gives a slightly larger result. For PbPb collisions, the normalization is enhanced by a factor of order $Z^2 A$, partially compensated by the smaller photon energies probed, which also reduce the BFKL enhancement relative to the Born term. In this case, the $Z^2$ factor corresponds to the nuclear photon flux, while the $A$ factor accounts for the nuclear PDFs. Predictions for Pb$p$ and $p$Pb collisions differ because the roles of photon emitter and target are reversed in each configuration. In Pb$p$ collisions, the nucleus acts as the photon emitter, enhancing the cross section by a factor of order $Z^2$. Conversely, in $p$Pb collisions, the nucleus serves as the hadronic target, enhancing the cross section by a factor of order $A$.

\begin{table*}[t]
\centering
\setlength{\tabcolsep}{8pt}
\renewcommand{\arraystretch}{1.30}
\begin{tabular}{lccccc}
\toprule
\textbf{System} & Born & LLA & NLA-BLM & NLA-All-poles & NLA-PMS \\
\midrule
$pp$  & 6.19 & 29.5 & 16.9 & 19.7 & 14.0 \\
Pb$p$ & $5.29\times10^{3}$ & $9.42\times10^{3}$ & $8.71\times10^{3}$ & $9.33\times10^{3}$ & $7.83\times10^{3}$ \\
$p$Pb & $3.43\times10^{2}$ & $1.36\times10^{3}$ & $9.08\times10^{2}$ & $1.04\times10^{3}$ & $7.63\times10^{2}$ \\
PbPb  & $1.19\times10^{6}$ & $1.70\times10^{6}$ & $1.69\times10^{6}$ & $1.76\times10^{6}$ & $1.56\times10^{6}$ \\
\bottomrule
\end{tabular}
\caption{Cross sections [nb] integrated over $1 < |t| < 100$~GeV$^2$ and over the photon rapidity, $-10 < y < 10$, with a rapidity gap $\Delta\eta > 3$ between the photon and the system $X$}
\label{tab:sigma}
\end{table*}

\begin{figure*}[h!]
\centering
\makebox[\textwidth][c]{%
\includegraphics[width=0.51\textwidth]{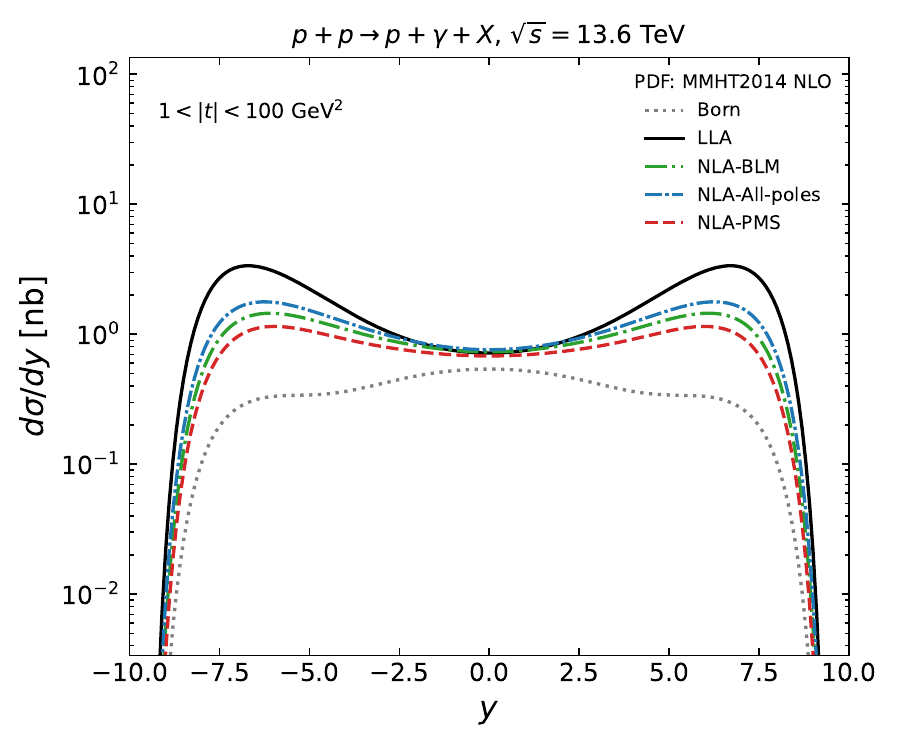}%
\hspace{-0.15cm}%
\includegraphics[width=0.51\textwidth]{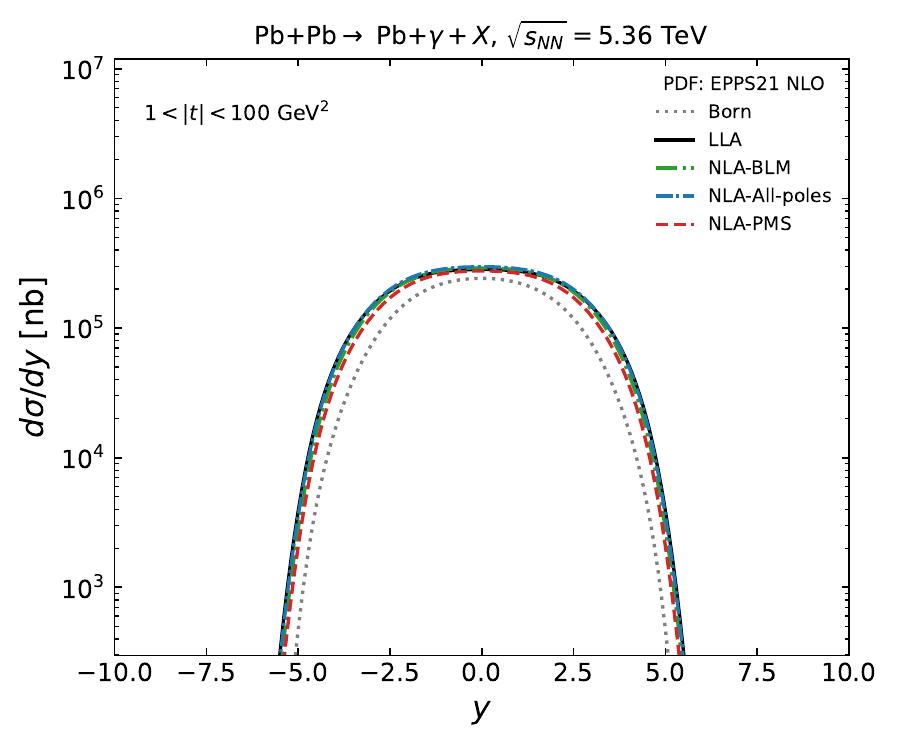}}
\\[0.20cm]
\makebox[\textwidth][c]{%
\includegraphics[width=0.51\textwidth]{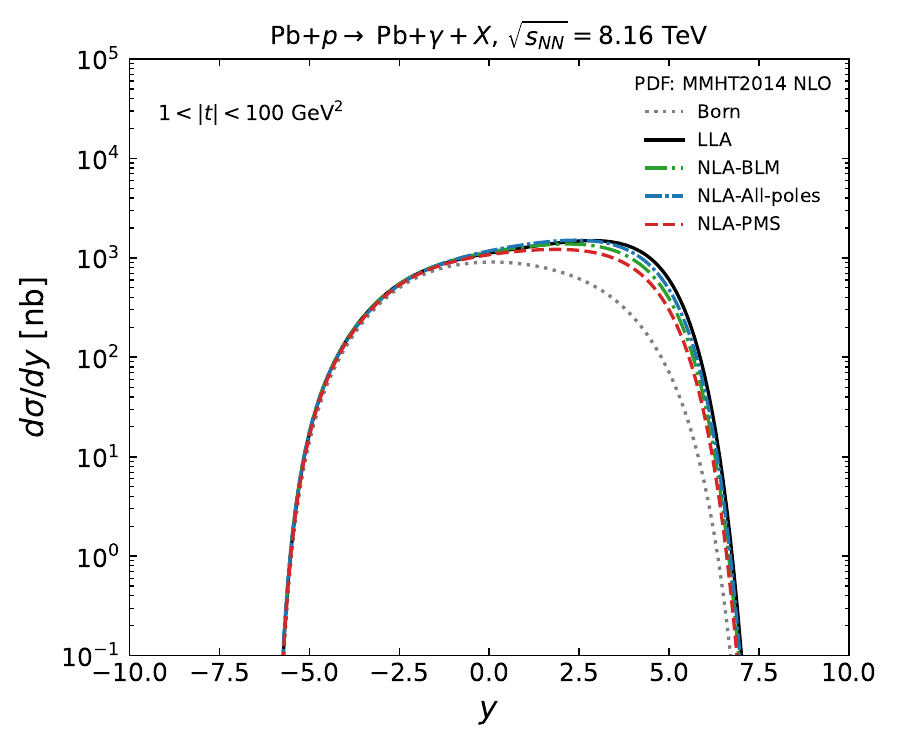}%
\hspace{-0.15cm}%
\includegraphics[width=0.51\textwidth]{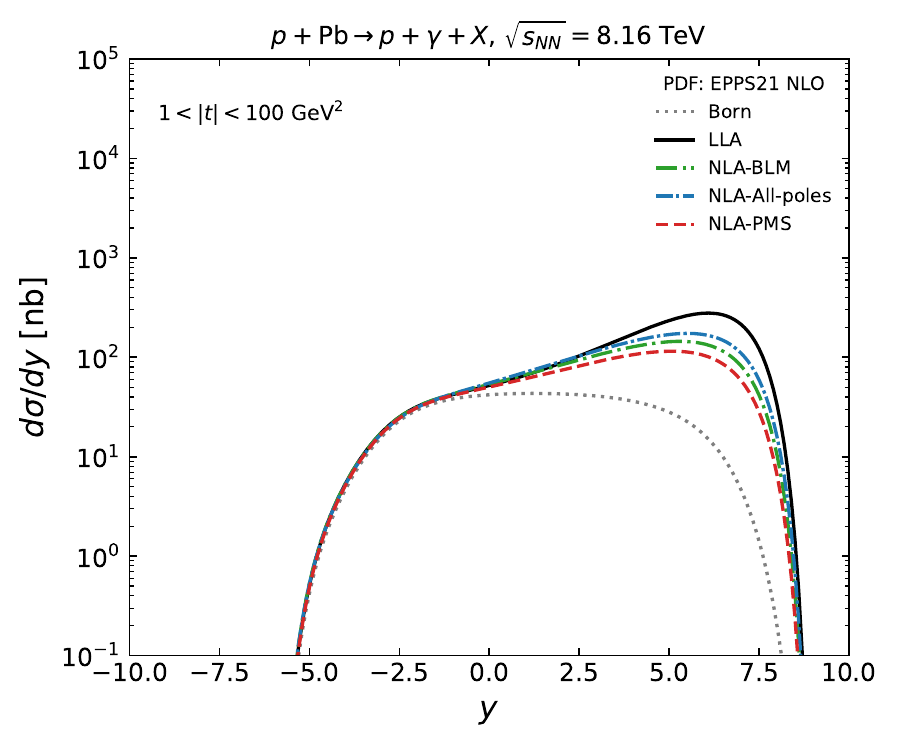}}
\caption{Rapidity distribution of the photon, $\drv\sigma/\drv y$, integrated over $1 < |t| < 100$~GeV$^2$, in $pp$ collisions at $\sqrt{s} = 13.6$~TeV (upper left), PbPb collisions at $\sqrt{s_{NN}} = 5.36$~TeV (upper right), and $p$Pb collisions at $\sqrt{s_{NN}} = 8.16$~TeV (lower panels). In the case of proton-lead collisions, we examine results considering either the nucleus (lower left) or the proton (lower right) as the source of photon emission.
The predictions with the Born term and the LLA, NLA-BLM, NLA-All-poles and NLA-PMS kernels are shown.}
\label{fig:dsdy}
\end{figure*}

In Fig.~\ref{fig:dsdy} we present our predictions for the rapidity distributions for diffractive photon photoproduction in $pp$ collisions at $\sqrt{s}=13.6$~TeV (upper left), PbPb collisions at $\sqrt{s_{NN}}=5.36$~TeV (upper right), and $p$Pb collisions at $\sqrt{s_{NN}}=8.16$~TeV (lower panels). The results are obtained by integrating over $1<|t|<100$~GeV$^2$ and by considering different models for the BFKL characteristic function. For $pp$ and PbPb collisions we predict symmetric distributions, as expected when summing the contributions in which either the forward- or backward-moving hadron acts as the photon emitter. Photon production in $pp$ collisions is more sensitive to QCD dynamics than in PbPb, as indicated by the larger enhancement of the BFKL predictions relative to the Born term at forward/backward rapidities; this follows from the fact that larger $|y|$ probes higher photon--hadron center-of-mass energies, which are intrinsically larger in $pp$ than in PbPb. For $p$Pb collisions, Fig.~\ref{fig:dsdy} shows the results obtained by taking either the nucleus (lower left) or the proton (lower right) as the photon emitter, leading to asymmetric rapidity distributions since we are considering that only one of the incident hadrons acts as the photon emitter. A larger enhancement is found when the proton emits the photon, albeit with a smaller overall normalization. These results indicate that a future analysis of this process with a forward detector---e.g. the proposed FoCal detector at ALICE~\cite{ALICE:2024jtt,Jonas:2024ftk,Park:2024lhr}, expected to cover $3.2<y<5.8$---will be able to probe the underlying QCD dynamics.

\begin{figure*}[h!]
\centering
\makebox[\textwidth][c]{%
\includegraphics[width=0.51\textwidth]{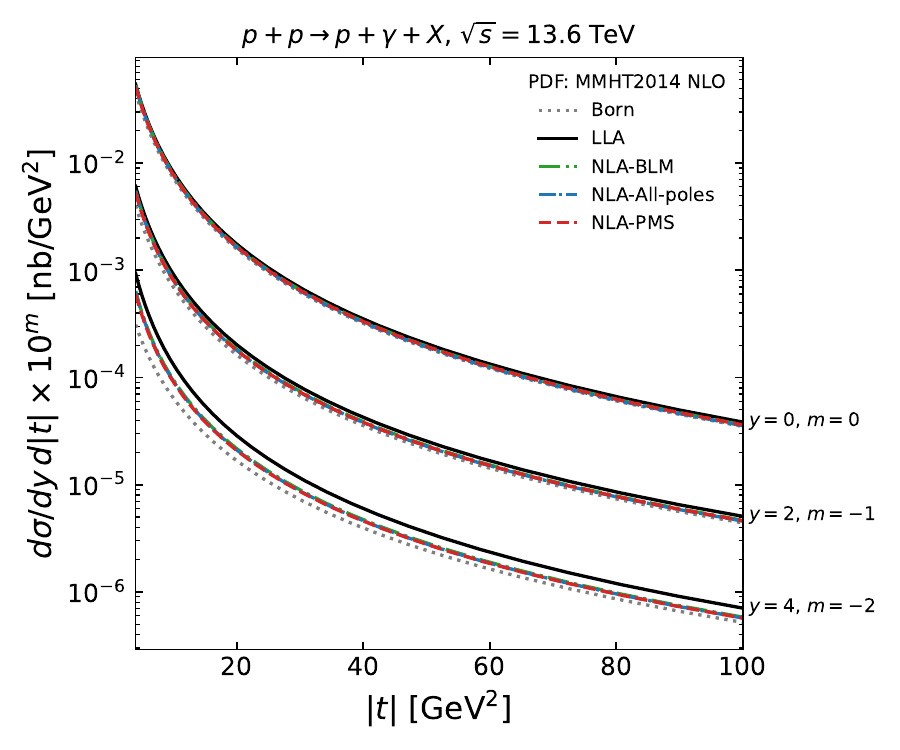}%
\hspace{-0.15cm}%
\includegraphics[width=0.51\textwidth]{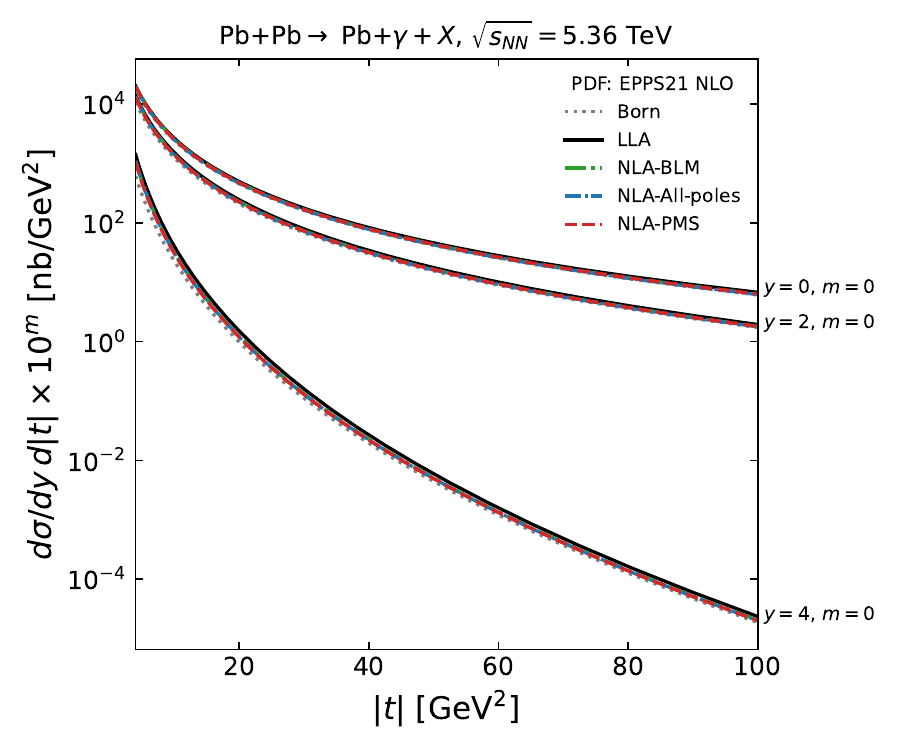}}
\\[0.20cm]
\makebox[\textwidth][c]{%
\includegraphics[width=0.51\textwidth]{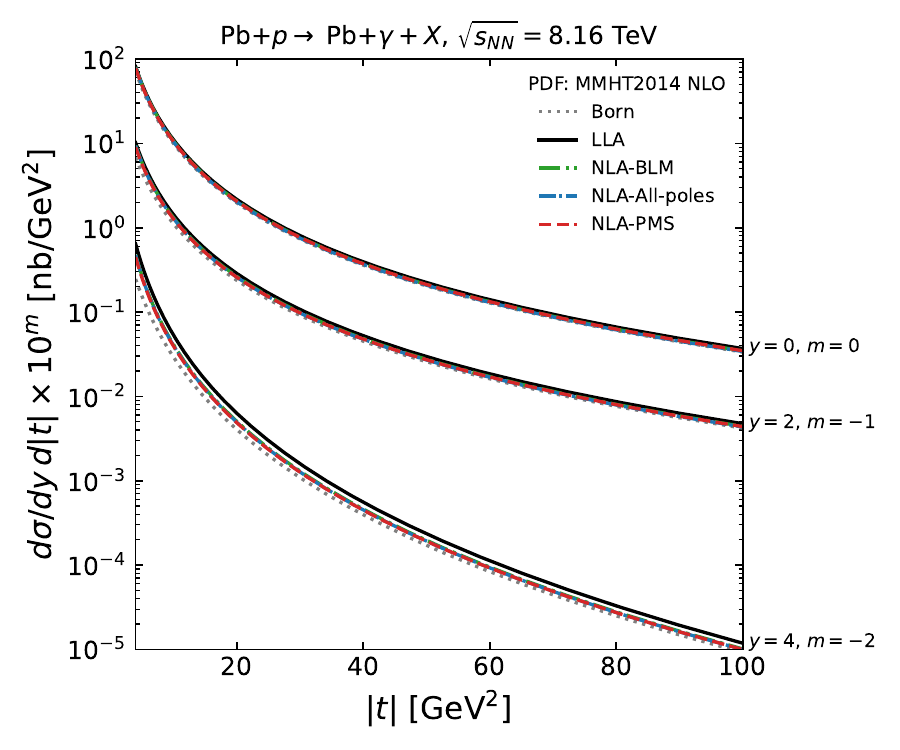}%
\hspace{-0.15cm}%
\includegraphics[width=0.51\textwidth]{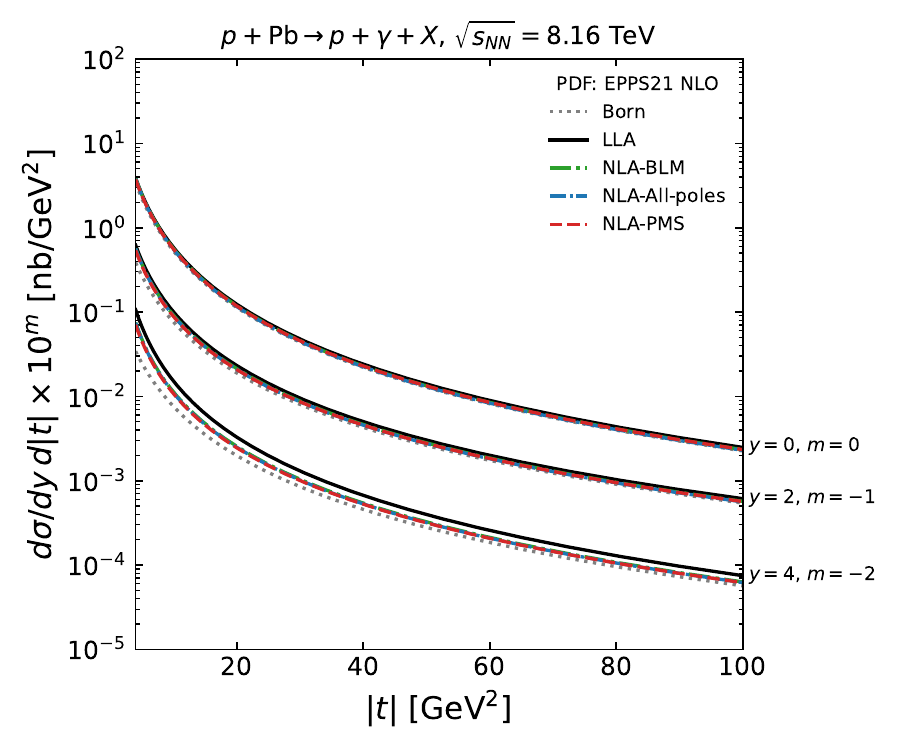}}
\caption{Double differential distribution $\drv\sigma/\drv y\drv|t|$ as a function of the squared transverse momentum of the produced photon ($|t| = p_T^2$) at fixed rapidities ($y=0, 2, 4$) for the colliding systems considered in Fig.~\ref{fig:dsdy}. Results for different rapidities have been rescaled by a multiplicative factor $10^m$ to avoid the overlap among  predictions.  }
\label{fig:dsdydt}
\end{figure*}

In Fig.~\ref{fig:dsdydt} we present our predictions for the double-differential distribution $\drv\sigma/\drv y\,\drv|t|$ as a function of the squared transverse momentum of the produced photon ($|t| = p_T^2$), for fixed rapidities ($y=0, 2, 4$). We consider the same colliding systems as in Fig.~\ref{fig:dsdy} and the results at different rapidities are rescaled by a factor $10^m$ to avoid overlap among the curves. As anticipated from Fig.~\ref{fig:dsdy}, the spread among the predictions grows toward forward rapidities and is more pronounced in $pp$ collisions. We find a steep decrease of the distribution with increasing $p_T$. Nevertheless, a non-negligible yield of photons with $p_T \approx 4$--$5$~GeV is expected in UPCs.

%-----------------------------------------
\section{Summary}
\label{sec:conclusions}
%-----------------------------------------

In this paper we have investigated, for the first time, diffractive photoproduction of high-$p_T$ photons in ultraperipheral collisions at the LHC. The theoretical description is closely related to that of large-$|t|$ diffractive vector-meson production, which has been extensively discussed in the literature. A key advantage of the photon final state is the absence of a vector-meson wave function, which simplifies the calculation. Apart from the BFKL energy scale, fixed by HERA data, the main non-perturbative input is provided by the proton parton distribution functions. The corresponding cross section is, however, suppressed relative to vector-meson production due to the electromagnetic coupling of the $q\bar{q}$ pair to the final-state photon. We have estimated the cross sections including NLO corrections to the BFKL characteristic function, and we have explored the prospects for measuring this process in ultraperipheral $pp$, $p$Pb, and PbPb collisions. We have also compared our framework with HERA data and derived parameter-free predictions for UPCs. Our results indicate that the predicted cross sections are non-negligible, and that future experimental studies of this channel can help constrain the BFKL dynamics and improve our understanding of QCD at high energies.

The results presented in this paper strongly motivate improving the theoretical description of the photon and quark impact factors by including next-to-leading-order corrections (see, e.g., Refs.~\cite{Chirilli:2013kca,Chirilli:2014dcb,Ivanov:2014hpa,Colferai:2023dcf}). Moreover, they motivate dedicated studies of the experimental feasibility of isolating this final state using realistic cuts and detector efficiencies. We intend to address both topics in forthcoming work.

%-----------------------------------------
\section*{Acknowledgments}
%-----------------------------------------
\label{sec:acknowledgments}
\addcontentsline{toc}{section}{\nameref{sec:acknowledgments}}
The authors acknowledge very useful discussions with David D'Enterria and Jeremi Niedziela, which motivated this study, as well as with Francesco Celiberto regarding higher-order effects in BFKL dynamics. VPG and JVB were partially supported by the Brazilian funding agencies CNPq, CAPES (Finance Code 001), FAPERGS, and INCT-FNA (Process No.~408419/2024-5).

%\clearpage
\bibliographystyle{apsrev}
\bibliography{bibliography}

\end{document}